\documentclass{raa}
\usepackage{graphicx,times}
\usepackage{natbib}
\usepackage{amssymb,amsmath}
\bibpunct{(}{)}{;}{a}{}{,}

\usepackage[a4paper=true,dvipdfm=true,pagebackref=true]{hyperref}
\hypersetup{pdftitle = The title of my PDF, pdfauthor = My name, pdfsubject= The subject, pdfkeywords = keyword1 keyword2 keyword3}
\hypersetup{colorlinks = true, linkcolor = green, anchorcolor = red, citecolor = blue, filecolor = red, pagecolor = red, urlcolor = red}

\begin{document}

   \title{Sunspots rotation and magnetic transients associated with flares in NOAA AR 11429
$^*$
\footnotetext{\small $*$ Supported by the National Natural Science Foundation of China.}
}

 \volnopage{ {\bf 2017} Vol.\ {\bf X} No. {\bf XX}, 000--000}
   \setcounter{page}{1}

   \author{Jianchuan Zheng\inst{1}
      \and Zhiliang Yang\inst{1}
      \and Jianpeng Guo\inst{1, 2}
      \and Kaiming Guo\inst{1}
      \and Hui Huang\inst{1}
      \and Xuan Song\inst{1}
      \and Weixing Wan\inst{2, 3}
   }

   \institute{ Department of Astronomy, Beijing Normal University,
               Beijing 100875,  China; {\it zhengjc@mail.bnu.edu.cn,
               jpguo@bnu.edu.cn}\\
        \and
              Key Laboratory of Earth and Planetary Physics, Institute
           of Geology and Geophysics, Chinese Academy of Sciences, Beijing
           100029, China\\
        \and
             College of Earth Sciences, University of Chinese Academy
             of Sciences, Beijing,
           China\\
               {\small Received -------; accepted ------}}


\abstract{
We analyze sunspots rotation and magnetic transients in NOAA AR 11429
during two X-class (X5.4 and X1.3) flares using the data from the Helioseismic
and Magnetic Imager on board the \emph{Solar Dynamics Observatory}.
A large leading sunspot with positive magnetic polarity
rotated counterclockwise. As expected, the rotation was significantly
affected by the two flares. The magnetic transients induced by the flares were
clearly evident in the sunspots with negative polarity. They were moving across
the sunspots with speed of order $3-7\ \rm km \ s^{-1}$. Furthermore,
the trend of magnetic flux evolution of these sunspots exhibited
changes associated with the flares.
These results may shed light on the understanding of the evolution of sunspots.
\keywords{Sunspots; Sun: rotation; Sun: magnetic fields
}
}

   \authorrunning{J.-C. Zheng et al. }            
   \titlerunning{Sunspots rotation and magnetic transients associated with flares}
   \maketitle

\section{Introduction}
\label{sect:intro}
Sunspots are concentrations of strong magnetic field on the solar surface,
consisting of a dark umbra and a fibrous penumbra \citep{sol2003}.
It is generally accepted that eruption events such as flares occurring in
upper atmosphere are associated with magnetic field generated from
the base of convection zone, emerging as bipolar magnetic regions at solar
surface, then extending to corona \citep{sol2006}. So the visible sunspots
play an important role in studying a variety of phenomena on the Sun.


The evolution of sunspots includes formation, movement, deformation and
disappearance. Rotation is an important feature during evolution.
In general, there are two kinds of rotation: one sunspot
rotates around its umbral centre and one sunspot rotates around the
other \citep{yan2012}.
The rotation of sunspots and spot-groups have been studied for many decades
(see, e.g., \citealt{eve1910, stj1913, mal1964, bro2003, zha2007, yan2009, min2009}).
\citet{yanx2008} found 182 significantly rotating
sunspots among 2959 active regions. Some authors believed the
rotational motion of sunspots may be involved with energy build-up and later
release by flares (e.g., \citealt{ste1969, bar1972, hir2005, zha2008}).
Furthermore, the relationships between the sunspot rotation and coronal
consequence \citep{bro2003, tia2008}, flare productivity
\citep{yan2008, zha2008, rua2014}, direction of the rotation \citep{zhe2016},
and magnetic helicities \citep{vem2012} have been investigated. The association
of flares with abnormal rotation rates \citep{hir2005} has been found.
Specifically, \cite{wan2014} reported two sunspots in AR 11158 rotating along
with a X2.2 flare. \cite{liu2016} reported a non-uniform rotation induced
by a M6.5 flare in NOAA AR 12371, and \cite{biy2016} found an abrupt reverse
rotation of sunspot in NOAA AR 12158 caused by back reaction of X1.6 flare.


Besides, the photospheric magnetic field
changes during the flares. The magnetic field is distorted enough to
store energy
powering flare. After a flare taking place, the distorted field relax and
restructure \citep{sak1996}. There are two kinds of changes of magnetic
field based on observations: the first is rapid short-term changes
\citep{pat1981, zha2009}, which is due to flare-induced spectral line changes
\citep{pat1984, qiu2003}. It is generally
believed that the rapid short-term changes are not the real changes of magnetic field (see,
\citealt{din2002}). However, \cite{har2013} also analyzed the magnetic
transients during the M7.9 flare in NOAA AR 11429 on 13 March 2012 and
suggested that the magnetic transients represented a real change in the
photospheric magnetic filed.
Another is  irreversible changes from pre-flare to post-flare
(e.g., \citealt{spi2002, wan2002, wan2006, son2016}).
The irreversible changes were first noticed by \cite{wan1994}.
They showed that magnetic shear increased after flares with vector magnetic
field data. \cite{cam1999} reported a change in the line-of-sight (LOS)
field during the X9.3 flare on 24 May 1990 with the videomagnetograph
data from the Big Bear Solar Observatory (BBSO). \cite{sud2005} and
\cite{pet2010} analyzed the changes of the LOS magnetic field during
some X- and M- class flares with Global Oscillation Network Group (GONG)
magnetograms. The median duration of changes was about 15 minutes and
the median absolute value of changes was about 69 G. \cite{wan2010}
demonstrated that the change of LOS magnetic field is due to increase
of horizontal magnetic field near the polarity inversion line.
With the seeing-free data of vector magnetograms, \cite{liu2012}
found that the photospheric transverse magnetic field enhancement was associated
with the M6.6 flare in NOAA AR 11158.
\cite{wan2012} observed the same phenomenon associated with the X2.2
flare in this active region. These results suggested that the magnetic field
near the polarity inversion line could become more horizontal after a flare.


In this study, we analyze the evolution of sunspots in NOAA AR 11429.
We will explore the rotation of sunspots along with the flares, i.e.,
how do the flares and the rotation of sunspots affect each other.
The magnetic transients induced by flares are studied in prior works.
We further study the motion of the magnetic transients to understand
how the flares influence sunspots. The
paper is organized as follows. In Section \ref{sect:Obs} we describe the
data used and the entire structure of the active region. Section
\ref{sec:analysis} shows data processing. We summarize and discuss the
results in Section \ref{sect:discussion}.

\section{Observational Data}
\label{sect:Obs}

The 45 s cadence and spatial sampling of $0''.5$ pixel$^{-1}$ full-disk
continuum intensity images and line-of-sight magnetograms observed by
the Helioseismic and Magnetic Imager (HMI; \citealt{sch2012}) on board
the \emph{Solar Dynamics Observatory} (\emph{ SDO}) are used to analyze
the variation of sunspots in NOAA AR 11429.
The images from the Atmospheric Imaging Assembly (AIA;
\citealt{lem2012}) on board \emph{SDO} are used to investigate the chromospheric
and coronal context: AIA 304 $\rm \AA$ formed in transition region and
chromosphere, and AIA 1600 $\rm \AA$ formed in upper photosphere and
transition region.

\begin{figure}
   \centering
  \includegraphics[width=14.5cm, angle=0]{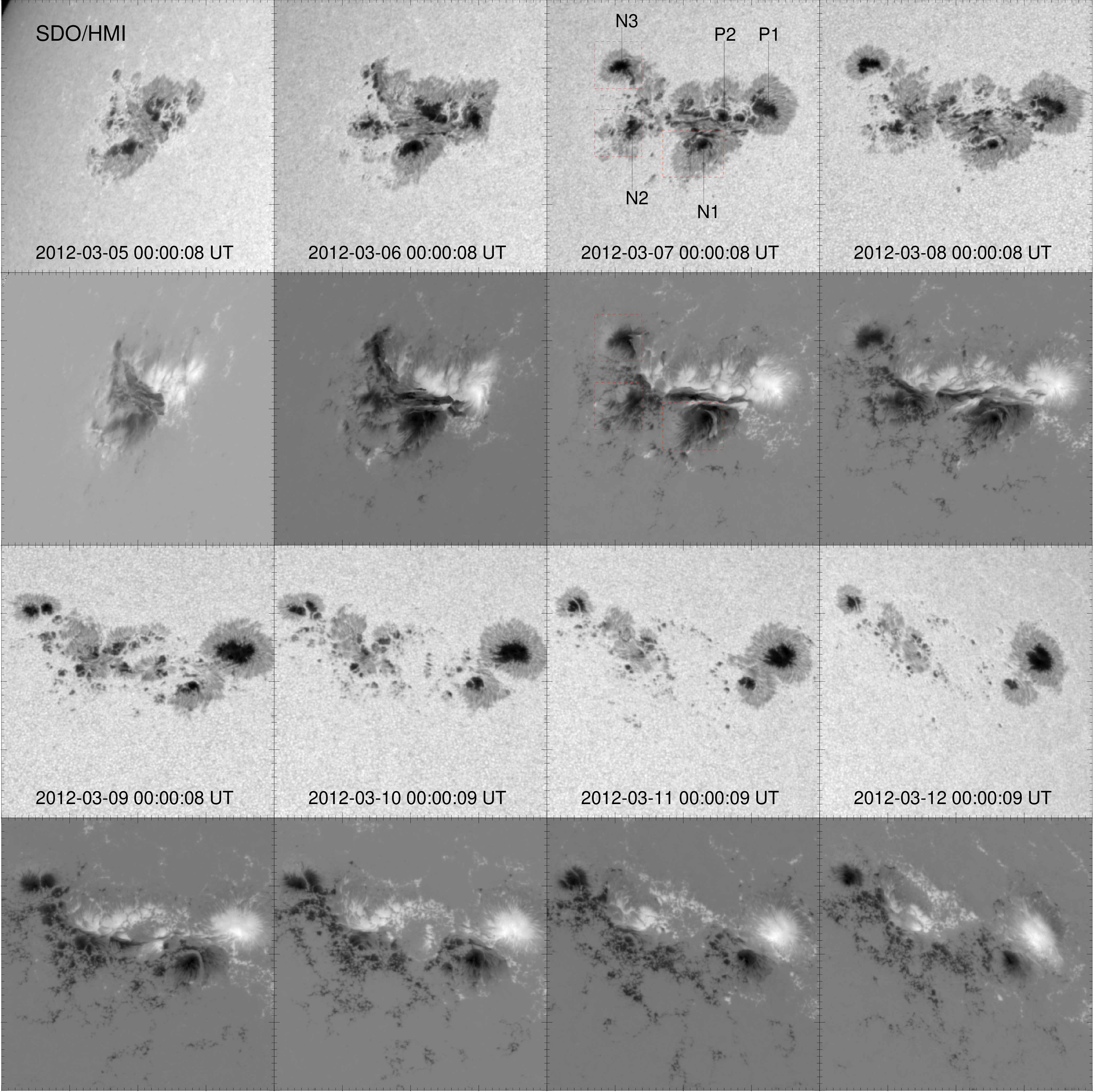}
   \caption{The HMI intensity images and magnetograms show the evolution
           of AR 11429 during 5 - 12 March 2012. The field of view is
           $200''\times200''$. The main sunspots are labeled with P/N*.
            The white/black areas in magnetograms represent
            positive/negative magnetic polarity.
           The red rectangular boxes are the selected regions for further
           study.}
   \label{fig1}
   \end{figure}

AR 11429 appeared on the eastern limb on 3 March 2012, and then rapidly became
complicated reverse-polarity $\beta\gamma\delta$ active region. The
evolution and development of AR 11429 from 5 March to 12 March are
displayed in Figure \ref{fig1}. It is interesting to note that the sunspots in this active
region were moving away from each other, i.e., the leading sunspots moving
westward and the following sunspots moving eastward. Several major
eruptions were clearly visible in this active region. Typically, three X-class
flares occurred: one (X1.1) on 5 March and two of interest (X5.4 and X1.3) on
7 March. The GOES soft X-ray flux (Figure \ref{fig3}a) indicates that
the X5.4 flare started at 00:02 UT, peaked at 00:27 UT, and ended at 00:40 UT,
and the X1.3 flare started at 01:05 UT, peaked at 01:14 UT, and ended at
01:23 UT. The evolution of five prominent
sunspots, two positive polarity sunspots (P1 and P2) and three negative
polarity sunspots (N1, N2 and N3), were associated with these two flares,
as be discussed in the next section.

\section{Analysis and Results}
\label{sec:analysis}

\subsection{Sunspots rotating associated with flares}
\begin{figure}
   \centering
  \includegraphics[width=14.5cm, angle=0]{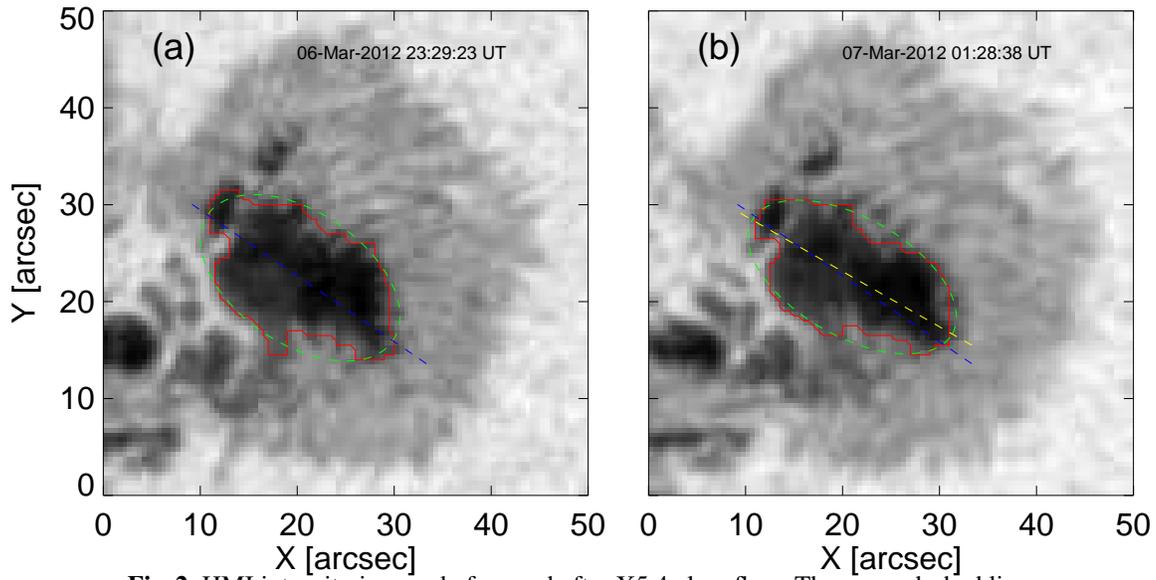}
   \caption{HMI intensity images before and after X5.4 class flare. The
            green dashed lines represent the ellipse fit to the sunspots
            P1. The blue and yellow dashed lines are the major axes of
            the ellipses.}
   \label{fig2}
   \end{figure}

\begin{figure}
   \centering
  \includegraphics[width=10cm, height=20cm, angle=0]{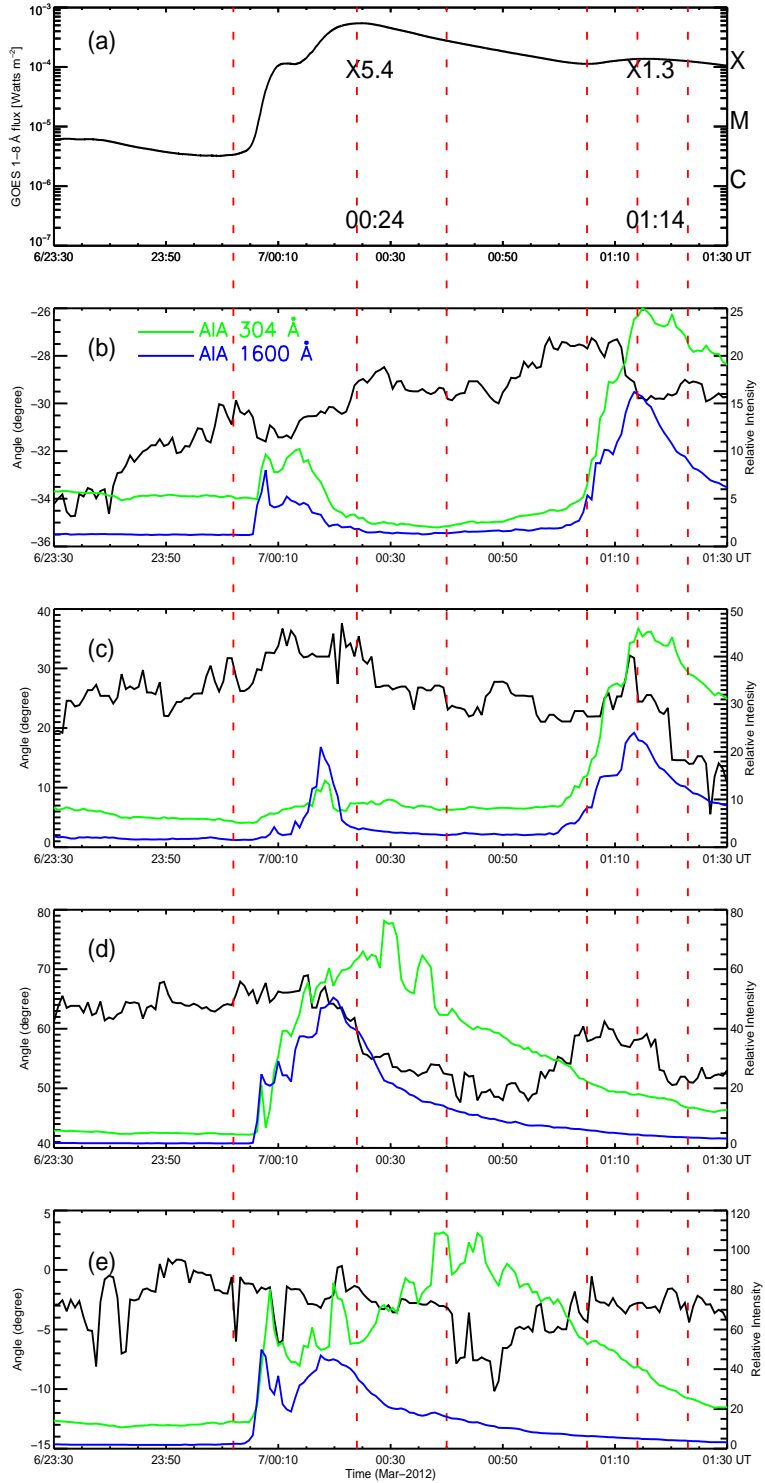}
   \caption{(a) Evolution of GOES soft X-ray emission from 23:30 UT on
            6 March to 01:30 UT on 7 March. The vertical dash lines
            represent the GOES flare start, peak and end times.
            (b)--(e) Time profiles of orientation angle
            (between the major axis and horizontal direction) of
            sunspot P1, N1, N2 and N3, respectively. The
            green and blue curves are profiles of relative mean
            intensity within each sunspot region with respect
            to the quiet Sun in AIA 304 $\rm \AA$ and AIA 1600
            $\rm \AA$ respectively.}
   \label{fig3}
   \end{figure}

To study the rotation, we take the rotation of individual sunspots as
a simplified solid-body rotation (i.e., a sunspot rotates around its
umbral center), and track the variation of the ellipses that best fit
individual sunspots on time-series intensity images (see \citealt{biy2016}). Take sunspot P1 as
an example. The best-fit ellipses on the intensity images at 23:29:23 UT
of 6 March and 01:28:38 UT of 7 March are shown in Figure \ref{fig2}(a)
and Figure \ref{fig2}(b), respectively. Obviously, the major
axis rotates counterclockwise about $7^{\circ}$. In the present study,
the changes of the angle between the major axis and the horizontal
direction of the best-fit ellipses are used to describe the rotation
of the sunspot. The results together with the GOES soft X-ray flux
are shown in Figure \ref{fig3}.


Figure \ref{fig3}(b) shows that
P1 rotated mainly counterclockwise from $\sim$23:30 UT of 6 March to
$\sim$01:30 UT of 7 March. Before the start of the X5.4
flare, the rotation was very fast with speed up to
$7.92^{\circ}\ \rm hr^{-1}$. From start to peak of the flare, the rotation
speed decreased slowly to about $4.89^{\circ} \ \rm hr^{-1}$ and the
rotation trend kept counterclockwise. Just after the peak of
the flare, the rotation trend reversed from counterclockwise
to clockwise. Its speed was about $4.07^{\circ} \ \rm hr^{-1}$.
After the flare, the rotation direction returned to counterclockwise
with the rotation speed of about $6.56^{\circ} \ \rm hr^{-1}$,
and the rotation speed
increased gradually until the beginning of the X1.3 flare. Later, the
rotation trend reversed to clockwise again
with the rotation speed of about $6.44^{\circ} \ \rm hr^{-1}$ and slowed down.
These results indicate that
the rotation of sunspot P1 was significantly affected by both flares.
The relative mean intensity profiles in different AIA wavelengths,
obtained from the co-aligned with HMI images, are also plotted in
Figure \ref{fig3}. The intensity in the AIA wavelengths of 304
$\rm \AA$ (green curve) and AIA 1600 $\rm \AA$ (blue curve) started
rising around 00:05 UT, peaked at about 00:07 UT, and then decreased
during the impulsive phase of the X5.4 flare. At about 01:05 UT, the
intensities started rising again and peaked at about 01:14 UT.
It provides further support for the view that solar flares are
often related to the rotation of sunspots.


The rotation of the sunspot N1 was complicated (Figure \ref{fig3}c). Before
the peak of the X5.4 flare, the rotation trend was mainly counterclockwise,
and then reversed to clockwise. The
situation around the X1.3 flare was similar to that around the former flare.
The intensity of AIA wavelengths increased and the
rotation changed significantly during the impulsive phases of the X5.4 and X1.3 flares.
As shown in Figure \ref{fig3}(d), sunspot N2 rotated clockwise from $\sim$00:15 UT to
$\sim$00:40 UT. The relative mean intensity of 1600 $\rm \AA$ in this sunspot
started rising at $\sim$00:05 UT and reached its maximum at $\sim$00:20 UT. The
relative mean intensity of 304 $\rm \AA$ started rising at $\sim$00:05 UT and reached its maximum
at $\sim$00:30 UT. So the X5.4 flare affected the rotation of sunspot N2.
The sunspot N3 remained stable during the two flares (Figure \ref{fig3}e).
 The rotation speed was poorly correlated
with the intensity of AIA wavelengths, and therefore the X5.4 and X1.3 flares.

\subsection{The motion of Magnetic Transient}
Magnetic transients induced by flares have been well investigated by
\cite{pat1984} and \cite{qiu2003}. They analyzed the cause, location and
level of the magnetic transients. Here, we find that there are some
magnetic transients moving through sunspots N1, N2 and N3,
and there is no apparent magnetic transient motion in
the sunspots P1 and P2. Because the polarity of
the sunspots N1, N2 and N3 is negative (black regions in the
magnetograms),
the magnetic transients means that the polarity is reversed.


Figure \ref{fig4} shows the changes of the magnetic field in sunspot N1.
At 01:09 UT of 7 March, the magnetic transients marked by red arrows in the umbra
of the sunspot can be clearly seen.
These magnetic transients (the positive magnetic field) moved toward the southeast
direction. At 01:14 UT, the area  with positive magnetic field became
larger while the strength became weaker, as shown in
Figure \ref{fig4}(a2).
The green line in Figure \ref{fig4}(a3)
indicates the moving trajectory of the positive polarity
points. To investigate the kinematic evolution of the reversed polarity,
the time-distance plots obtained along the green trajectory line are shown
in Figure \ref{fig4}(d). A white streak is clearly visible.
It started at $\sim$01:07 UT and ended at $\sim$01:16 UT,
 consistent with
the time of the X1.3 flare. It might imply that the reversed magnetic polarity
was caused by this flare. The distance corresponding to the streak  was about
5 arcsec. Thus, the moving speed of the magnetic transients can be
estimated to be about $6.59\ \rm km\  s^{-1}$.
Similarly, some magnetic transients (i.e., the reversed
magnetic field) were presented
in sunspots N2 and N3, and probably associated with the X5.4 flare, as shown
in Figures \ref{fig5} and \ref{fig6}. They were moving toward southeast
with speed about $3.69\ \rm km\  s^{-1}$ and $3.22\ \rm km\  s^{-1}$ respectively.

In order to check whether the magnetic transients were co-spatial
with flare ribbons, we further examined the co-aligned AIA 304 $\rm \AA$
and AIA 1600 $\rm \AA$ images corresponding to HMI magnetograms, which are
displayed in panels (b1)--(b3) and panels (c1)--c(3) in Figures \ref{fig4},
\ref{fig5} and \ref{fig6} respectively.
The flare ribbons can be seen in these panels. They were
separating from each other. The locations of sunspots N1, N2 and N3 in
AIA images, marked by magnetogram contours at -800 G,
were moving with flare ribbons. This implies the magnetic transients were
co-spatial with flare ribbons.

We calculated magnetic flux of sunspots N1, N2 and N3 for the regions
marked by the red rectangular boxes in Figure \ref{fig1}.
The magnetic flux of sunspots with negative polarity together with GOES
soft X-ray flux are shown in Figure \ref{fig7}. Before the start
of the X5.4 flare, the absolute value of magnetic flux in sunspot N1 decreased
slowly. From the start to peak of the flare, it decreased gradually. Then,
the value kept almost unchanged. At the end of the flare, it increased slightly. During the
period of the X1.3 flare, the absolute value of magnetic flux decreased first and then
increased abruptly. These results suggest that the changes of trend
of magnetic flux in sunspot N1 were caused
by both the X5.4 flare and the X1.3 flare.


The magnetic flux of sunspot N2 mainly increased from 23:30 UT of 6 March
to 01:30 UT of 7 March. There was an obvious change near the peak of X5.4 flare. Specifically,
the absolute value decreased first and then increased, as shown in Figure \ref{fig7}(c).
The absolute magnetic flux value of sunspot N3 (Figure \ref{fig7}d)
almost kept unchanged before the start of the X5.4 flare. Similarly, during
the X5.4 flare , the magnetic flux decreased rapidly, and then increased monotonically.
These results suggest that the changes of trend of magnetic flux
in sunspots N2 and N3 were mainly due to the X5.4 flare.

\section{Summary and Discussion}
\label{sect:discussion}

AR 11429 is a complicated active region with a reversed polarity structure,
in contrast with the nature of solar cycle 24. It has a large leading
sunspot with positive magnetic polarity (P1) exhibiting significant
counterclockwise rotation. The rotation underwent some changes during two
X-class (X5.4 and X1.3) flares. It is possible that the flares were
triggered by the sunspot rotation (see \citealt{zha2008, vem2016}),
and in turn they reacted back and affected the sunspot P1 rotation.

\begin{figure}
   \centering
  \includegraphics[width=6.5in, angle=90]{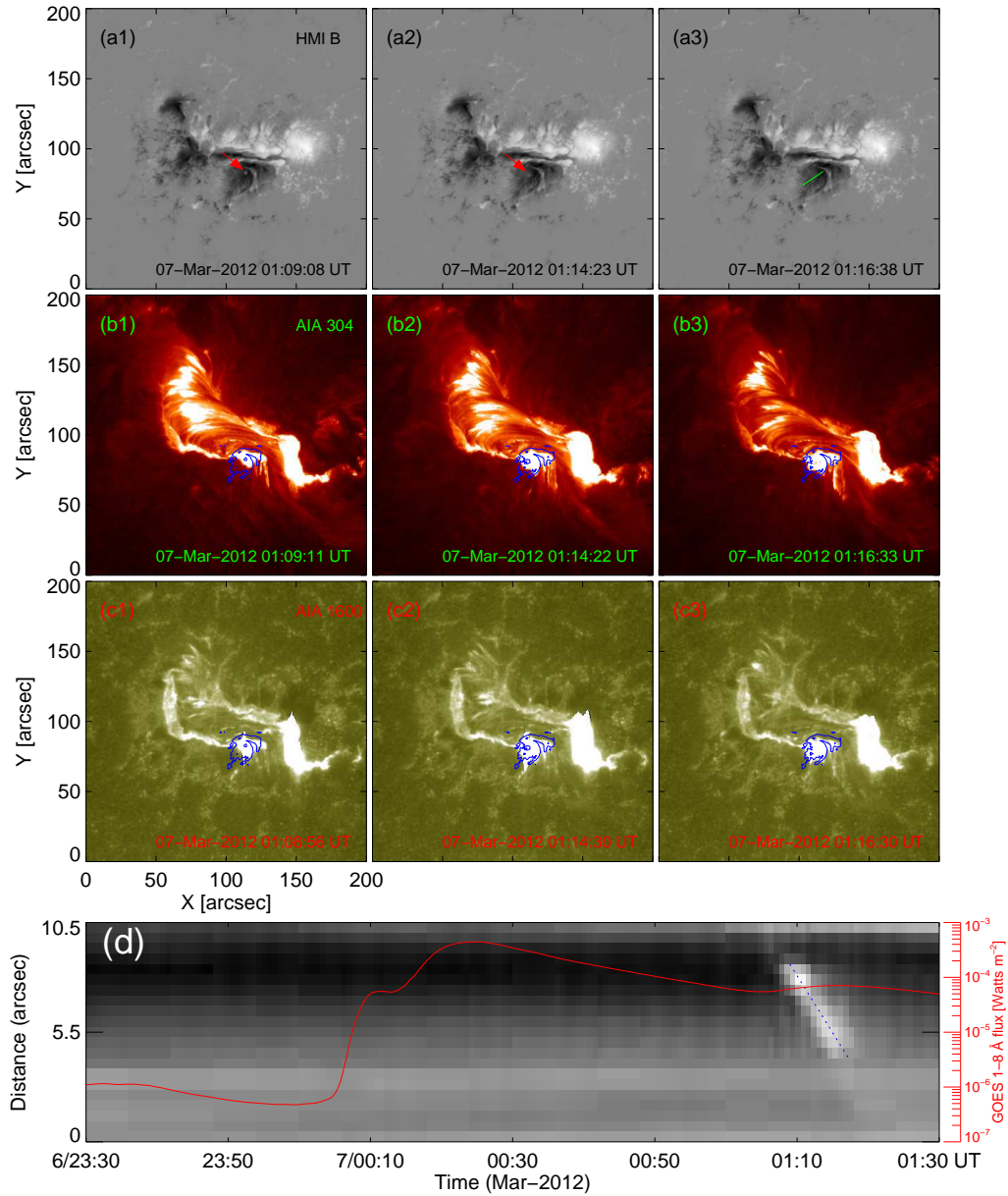}
   \caption{The reversed magnetic field moving through sunspot N1. The
            red
            arrows in (a1) and (a2) indicate the location of the reversed
            magnetic field and the green line in (a3) indicates the
            trajectory of the moving reversed magnetic field.
            (b1)--(b3) The co-aligned AIA 304 $\rm \AA$ images
            corresponding to HMI magnetograms. The blue contours represent
            the magnetic field of sunspot N1 at -800 G. (c1)--(c3) The
            co-aligned AIA 1600 $\rm \AA$ images overlaid by magnetogram
            contours of sunspot N1. (d) Time-slice
            plot of the trajectory. The blue dashed line shows the evolution
            of the reversed magnetic field. The red curve overlaid on the
            time-slice plot is the GOES soft X-ray flux.}
   \label{fig4}
   \end{figure}
\begin{figure}
   \centering
  \includegraphics[width=6.5in, angle=90]{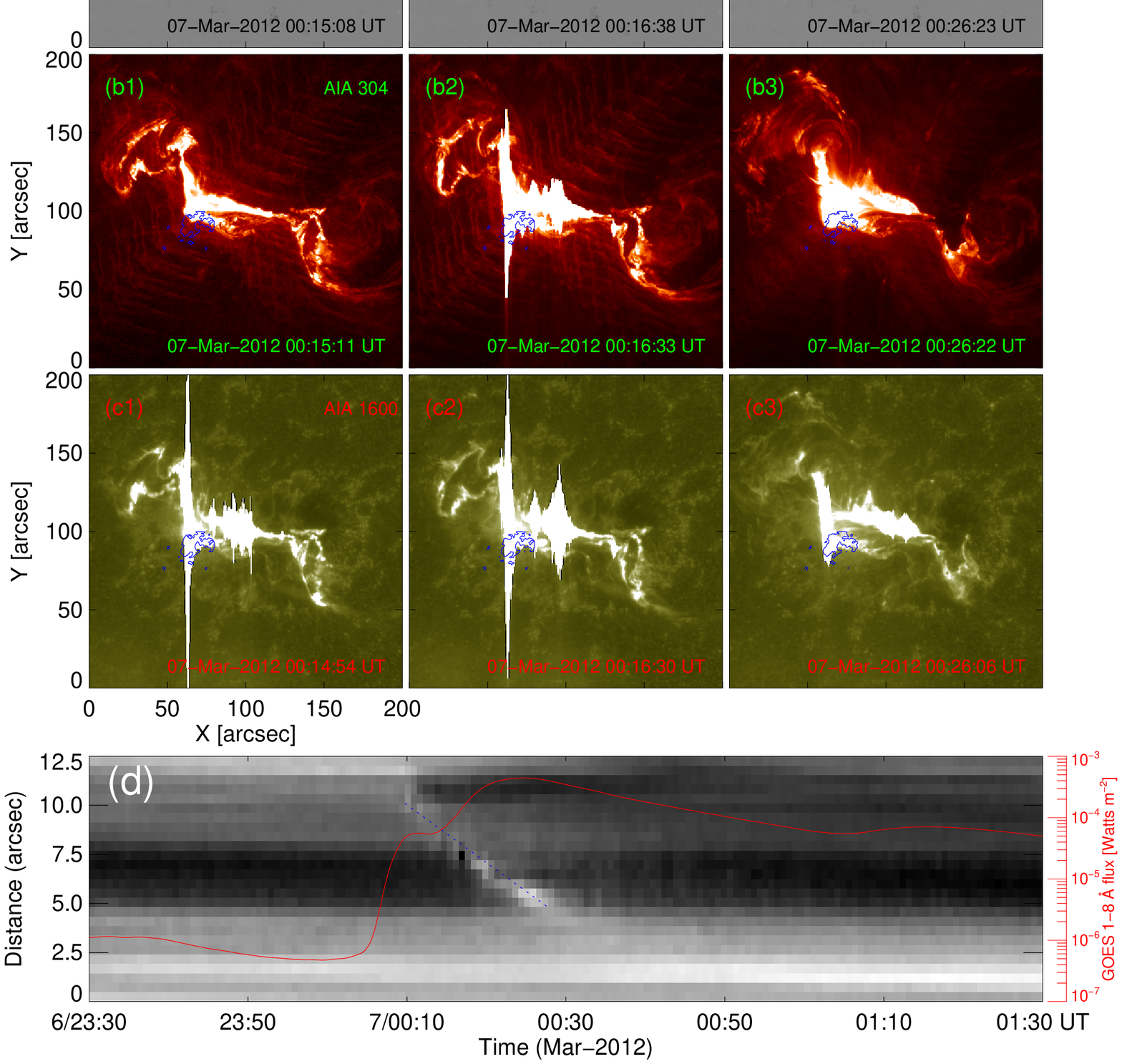}
   \caption{Same as Figure \ref{fig4}, but for sunspot N2.}
   \label{fig5}
   \end{figure}
\begin{figure}
   \centering
  \includegraphics[width=6.5in, angle=90]{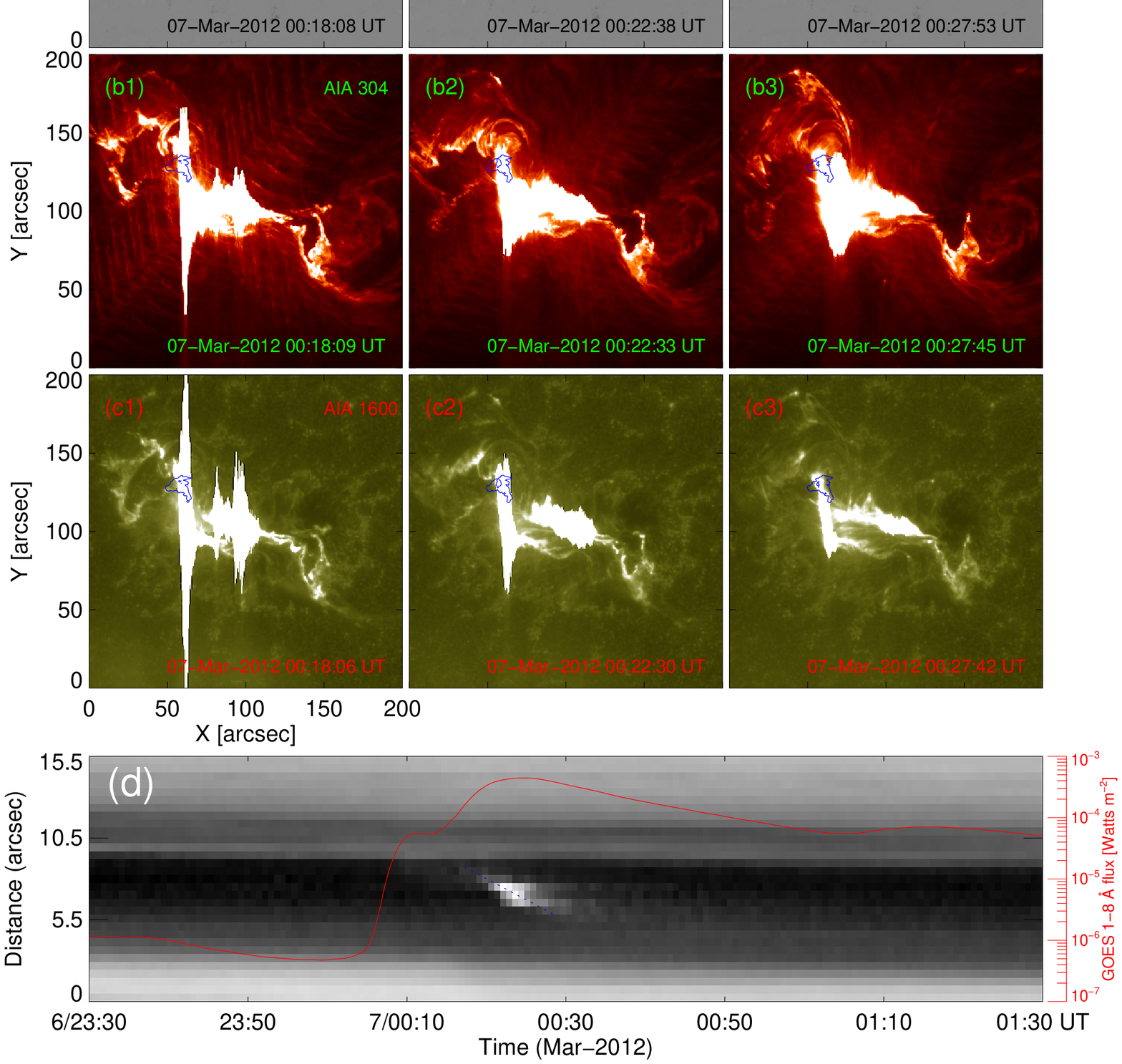}
   \caption{Same as Figure \ref{fig4}, but for sunspot N3.}
   \label{fig6}
   \end{figure}
\begin{figure}
   \centering
  \includegraphics[width=14.5cm, angle=0]{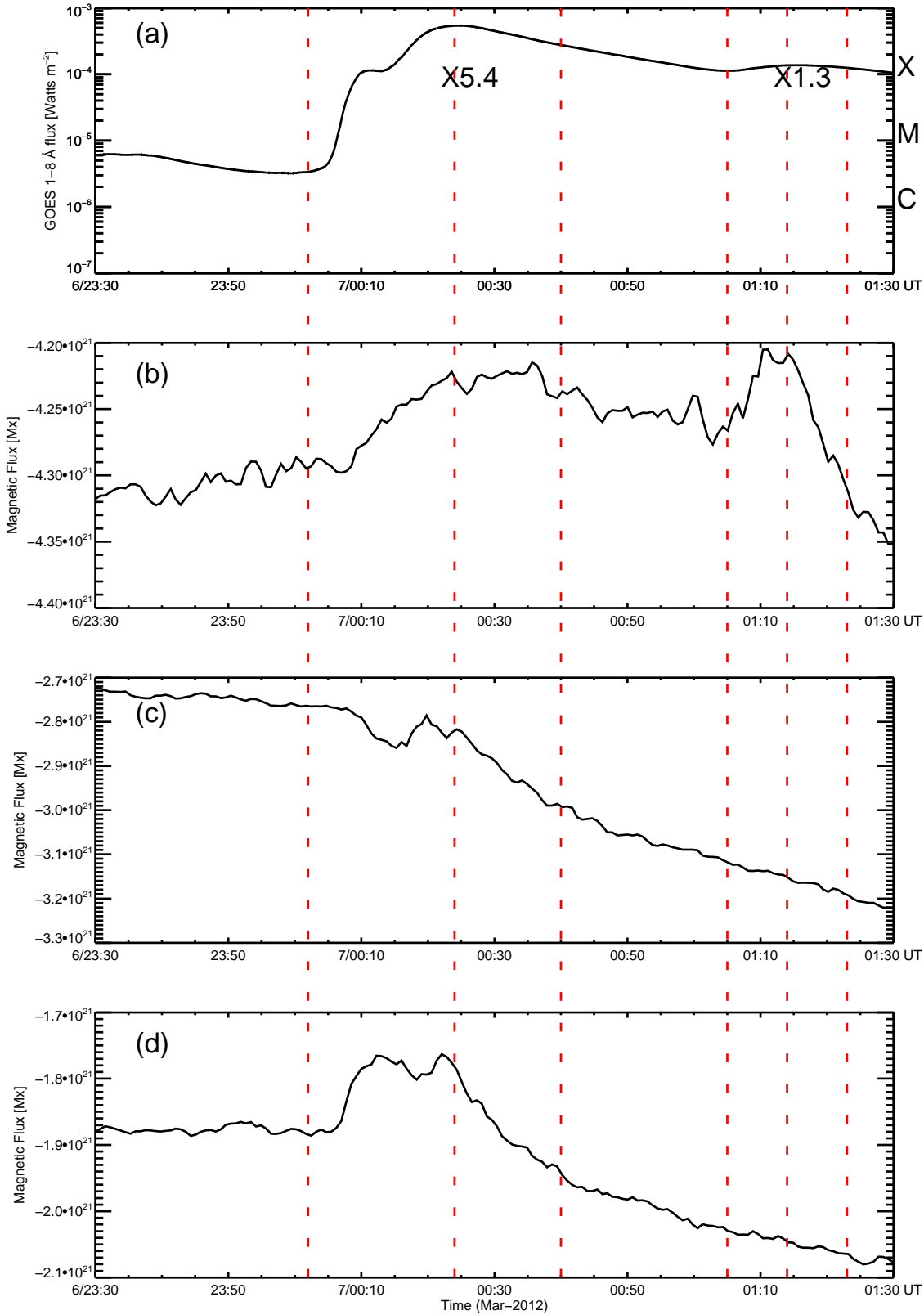}
   \caption{Magnetic flux of sunspot N1, N2 and N3.
            (a) Evolution of GOES soft X-ray emission from 23:30 UT on
            6 March to 01:30 UT on 7 March. The vertical dash lines
            represent the GOES flare start, peak and end times. (b)--(d)
            Time profiles of magnetic flux of sunspot N1, N2 and N3,
            respectively.}
   \label{fig7}
   \end{figure}

The magnetic transients induced by the two X-class flares can be seen in
sunspots with negative magnetic polarity (N1, N2 and N3). They moved outward
with respect to the center of the active region and moved through the
sunspots quickly with speed up to about $6.59\ \rm km \ s^{-1}$.
The magnetic transients lasted for about ten minutes within the period of
the X5.4 and X1.3 flare. As expected, the magnetic
transients, which are clearly illustrated in sunspots N1, N2 and N3 were moving away from the polarity
inversion line shown in Figures \ref{fig4}, \ref{fig5} and
\ref{fig6}. The magnetic transients occurred during the
X5.4 and X1.3 flares were co-spatial with flare ribbons. This is
consistent with \cite{mau2012}, who found that
magnetic transients in AR 11158 during a X2.2 flare persisted for
a few minutes and showed co-spatial with flare ribbons, which were
separating out with a mean velocity of 8 $\rm km\ s^{-1}$.

During the evolution of sunspots (N1, N2 and N3), changes in the
trend of magnetic field evolution were presented, and are associated with
the two X-class flares. Similar results in five $\delta$ sunspots were found
by \cite{wan2006}. The increasing and decreasing tendency of the magnetic
flux in active region might be attributed to magnetic flux emergence
and magnetic cancellation \citep{van2015}.


The evolution of sunspots with complicated structure can be affected by many
factors such as interaction among sunspots within same active region,
physical change induced by flare, and interaction between magnetic field
and plasma. In this study, we have found that the motion and
the trend of magnetic field evolution of sunspots
are significantly affected by flares. Other influences on the evolution
of sunspots will be examined in future studies.

\normalem
\begin{acknowledgements}
The authors thank the referees for their corrections
and valuable suggestions to improve the paper
This work is supported by the National Natural Science Foundations of
China (41231068, 41374187, 41531073, and 41674147). We acknowledge the
\emph{SDO}/HMI and \emph{SDO}/AIA teams
for providing the data. The GOES soft X-ray flux data is available at
NASA/GSFC Solar Data Analysis Center (SDAC).
\end{acknowledgements}

\bibliographystyle{raa}
\bibliography{ref}

\newpage

\end{document}